\documentclass[onecolumn,prd,amsmath,amssymb,amsfonts]{revtex4-2}

\usepackage{amsmath,amssymb,bm}

\usepackage{graphicx}
\usepackage{graphics}
\usepackage{dcolumn}
\usepackage{hyperref}
\usepackage{xcolor}
\usepackage[utf8]{inputenc}

\def\BEq{\begin{equation}}
\def\EEq{\end{equation}}
\def\BEqA{\begin{eqnarray}}
\def\EEqA{\end{eqnarray}}
\def\BEn{\begin{enumerate}}
\def\EEn{\end{enumerate}}
\def\BWT{\begin{widetext}}
\def\EWT{\end{widetext}}

\def\a{\alpha}

\def\b{\beta}

\def\L{\Lambda}

\usepackage{natbib}
\usepackage{cleveref}

\newcommand{\lr}[1]{\left( #1 \right)}
\newcommand{\R}{R_{\mu\nu}}
\newcommand{\La}{\Lambda}
\newcommand{\g}{g_{\mu\nu}}
\newcommand{\de}{\nabla_\mu\nabla_\nu}

\begin{document}

\title{Emergent vacua and stability constraints on black hole 
solutions in higher-dimensional \texorpdfstring{$f(R)$}{f(R)} gravity}

\author{Nicol\'{a}s Trullols Sandino and Andrei Galiautdinov}
\affiliation{Department of Physics and Astronomy, University of Georgia, Athens, Georgia 30602, USA}

\date{\today}

\begin{abstract}
We investigate static spherically symmetric vacuum solutions 
in higher-dimensional $f(R)$ gravity, beginning with the five-dimensional 
Starobinsky model governed by the action $f(R) = R + \alpha R^2 - 2\Lambda$. 
By enforcing the ghost-free stability criterion $f'(R) > 0$ on constant scalar 
curvature spacetimes, we show that a stable effective cosmological constant 
cannot be dynamically generated from pure $R^2$ geometric corrections 
in five dimensions; its existence is inextricably tied to a bare cosmological 
constant. Generalizing this analysis to arbitrary dimensions $D$ and 
single-term curvature corrections $f(R) = R + \alpha R^n$ with a vanishing 
bare cosmological constant, we derive a universal stability bound, $n > D/2$, 
required for the existence of stable emergent vacua. Finally, we demonstrate 
that expanding the gravitational action to a multi-term polynomial hierarchy 
circumvents this strict limitation. By including curvature corrections up to 
$\mathcal{O}(R^3)$, the extended geometric degrees of freedom simultaneously 
satisfy the trace constraint and the stability criterion. Furthermore, we establish 
the exact parameter space boundaries that ensure not only asymptotic 
vacuum stability but strict global stability ($f'(R) > 0$ for all $R$), allowing 
for the dynamical generation of exact, globally ghost-free vacuum spacetimes 
in $D \ge 5$ purely from higher-order geometric terms. The scalaron mass requirement, 
$m_s^2>0$ is not imposed in full in our analysis. A detailed investigation of its 
implications on the models considered in this work are left for future study.
\end{abstract}
\maketitle

\section{Introduction}

Modified theories of gravity have been an active area of research in an attempt 
to address the ultraviolet and infrared limitations of General Relativity. Among 
these theories, $f(R)$ theories of gravity compose a simple and well-developed 
subset of models, which extend the Einstein-Hilbert action by introducing an 
arbitrary function of the Ricci scalar $R$ 
\cite{De_Felice_2010,Sotiriou_2010,Capozziello2008,Capozziello_2011,PhysRevD.32.2511,NOJIRI_2007,Nojiri_2011,Nojiri_2017,munozpalma2026scalaronmodifiednullfocusingradial,
Panah_2026, verma2026cosmicstructureformationviable, rksz-2qzk,202605.0302}. 
Quadratic curvature corrections serve as a well-motivated geometric extension 
and have been extensively investigated in the context of quantum gravity and 
cosmology \cite{Stelle_1977}. In particular, the Starobinsky model, $f(R) = R +\a R^2$, 
has been a foundational example; it naturally emerges from leading-order quantum 
field corrections to the gravitation action and provides a standard approach to 
primordial inflation \cite{Starobinsky_1980,starobinsky2007disappearing,buchmuller2013starobinsky,
ivanov2022analytic,broy2016starobinsky,da2021brane}.
Additionally, the investigation of gravity in higher dimensions remains 
an important theoretical pursuit \cite{popov2026blackholesfrtheory, hell2026deal,anastasi2025relative,frassino2023higher, bueno2025regular, troncoso2000higher,charmousis2009higher,do2016higher,mignemi1992black}, 
motivated by braneworld scenarios \cite{Randall_1999}, Kaluza-Klein compactifications \cite{abe2026conjugateboundaryconditionskaluzaklein, karmakar2026kaluzakleingravitonshighercurvature,Bailin_1987, overduin1997kaluza,duff2025kaluzakleinsupergravity2025,chopovsky2012weak,eingorn2013kaluza,ali2026infrared,hosseinifar2026rotating,wesson1999space,reddy2012kaluza}, and string theory \cite{becker2007string,ibanez1945string,cicoli2026recent,wu2026revisiting,hao2026charged}. Exploring exact solutions within higher-dimensional modified theories provides 
a controlled environment to study the relationship between extended geometric 
degrees of freedom and the underlying structure of the vacuum \cite{de_la_Cruz_Dombriz_2009,ishibashi2003stability,coley1994higher,an2018examples,figueras2008extremal}.

The addition of higher-curvature terms to the action generates highly nonlinear 
field equations. To find exact spherically symmetric solutions such 
as higher-dimensional Tangherlini black holes, it is standard practice to 
impose a scalar curvature Ansatz \cite{Multam_ki_2006,Yu_2018,Podolsk__2020,Caram_s_2009}. 
This approach reduces the differential field equations to an algebraic constraint. 
Historically, this algebraic reduction has been used to demonstrate that 
$f(R)$ models can admit effective Einstein spaces \cite{vacaru2014ghost} 
characterized by an effective cosmological constant, $\L_{\text{eff}}$, generated 
dynamically by the curvature corrections themselves, even in the absence of 
a bare cosmological constant.

However, the mathematical existence of these constant-curvature geometries 
must be validated against physical stability conditions. In $f(R)$ gravity, 
the derivative of the gravitational Lagrangian with respect to the Ricci scalar, 
$f'(R)$, controls the sign of the effective gravitational coupling. If $f'(R)<0$, 
the effective Newton's constant becomes negative, introducing ghost instabilities 
characterized by propagating degrees of freedom with negative kinetic energy. 
Consequently, physically viable solutions are required to satisfy 
$f'(R) > 0$ \cite{Cognola_2006,cognola2010}. 

In this work, we analyze the ability of $f(R)$ gravity to generate stable 
emergent vacua. We begin in \ref{sec:5dStarobinsky} by revisiting static, 
spherically symmetric vacuum solutions in the five-dimensional Starobinsky 
model augmented by a bare cosmological constant, $\La$. Applying the stability 
criterion, we map the branch structure of the allowed spacetimes and demonstrate 
that pure $R^2$ corrections fail to generate a stable geometry on their own. 
In \ref{sec:singleTermCorrection}, we generalize this failure to arbitrary 
spacetime dimensions $D$ and single-term curvature corrections 
$f(R) = R+\a R^n$. We derive a universal stability bound, demonstrating 
that purely geometric, ghost-free vacua require $n > D/2$. Finally, 
in \ref{sec:multiTermCorrection}, we show that this restrictive bound can 
be circumvented by expanding the action to a multi-term polynomial 
effective field theory. We demonstrate that incorporating higher-order 
terms, such as $R^3$, provides the necessary degrees of freedom to 
dynamically generate stable, exact vacuum spacetimes across arbitrary 
dimensions without requiring a bare cosmological constant. 

\section{The five-dimensional Starobinsky model}
\label{sec:5dStarobinsky} 
\subsection{The field equations}

We begin with the five-dimensional Starobinsky model described by 
the Einstein-Hilbert action, augmented by a bare cosmological constant $\La$. 
The action is given as a function of the metric by,
\begin{equation}
S[g] = \frac{1}{16\pi G} \int d^5x\sqrt{-g}(R+\alpha R^2-2\La),
\label{Eq:action}
\end{equation}

where $\alpha$ is the Starobinsky parameter regulating 
the strength of the quadratic curvature correction. We utilize an extended 
form of the conventional metric signature $(-++++)$. Varying this action 
with respect to the metric tensor we get the $\La$-vacuum field equations, 
\begin{equation}
(\R-\frac{1}{2}\g R) +2\alpha R (\R -\frac{1}{4}\g R)-2\alpha(\de R - \g \Box R)+\g \La = 0,
\label{Eq:Fieldeqns}
\end{equation}
where $\Box  = g^{\alpha\beta}\nabla_\alpha\nabla_\beta$ is the d'Alembert operator. 
Taking the trace of Eq.\ (\ref{Eq:Fieldeqns}) provides the scalar constraint, 
\begin{equation}
8\alpha \Box R = \frac{3}{2}R+\frac{1}{2}\alpha R^2 -5\La
\label{Eq:scalarconstraint}
\end{equation}

Substituting this trace relation back into the original field 
equations allows us to eliminate the $\Box R$ term, producing a form 
that is more amenable to work with, 
\begin{equation}
(\R -\frac{1}{8}\g R) +2\alpha R (\R-\frac{3}{16}\g R)-2\alpha \de R-\frac{1}{4}\g \La = 0
\label{Eq:subbedtrace}
\end{equation}
Assuming constant Ricci curvature, $R=A=\text{constant}$, the full field 
equations reduce to the form,
\begin{equation}
f'(A)(\R -\frac{A}{5}\g) = 0,
\label{Eq:reducedFE}
\end{equation}
with $f'(A) = 1+2\alpha A$. 
Additionally, Eq.\ (\ref{Eq:scalarconstraint}) gives,
\begin{equation}
\alpha A^2 +3 A -10\La = 0, 
\label{Eq:scalareqn}
\end{equation}
or,
\begin{equation}
A = \frac{-3\pm\sqrt{9+40\alpha\La}}{2\alpha},
\label{Eq:Asol}
\end{equation}
with the bare cosmological constant, $\Lambda$, \emph{formally} restricted to
\begin{equation}
\Lambda \geq -\frac{9}{40 \alpha}.
\end{equation}
In order to avoid the ghost instability we further restrict $\Lambda$ by requiring 
\begin{equation}
f'(A) = -2\pm \sqrt{9+40\alpha\La} \geq 0,
\end{equation}
which is only possible for the ``positive'' branch,
\begin{equation}
f'(A) = -2 + \sqrt{9+40\alpha\La} \geq 0.
\end{equation}
This eliminates the ``negative'' branch, resulting in the condition,
\begin{equation}
\Lambda \geq -\frac{1}{8 \alpha},
\quad
A = \frac{-3 + \sqrt{9+40\alpha\La}}{2\alpha} \geq -\frac{1}{2\alpha}.
\end{equation}

\subsection{Exact Solutions}
 To search for black hole solutions, we impose a five-dimensional, static, 
 spherically-symmetric metric Ansatz, with a metric function $h(r)$,
\begin{equation}
ds^2 = -h(r)dt^2 + \frac{dr^2}{h(r)}+r^2d\Omega^2_3,
\label{Eq:ansatz}
\end{equation} 
where $d\Omega^2_3$ is the metric on the unit three-sphere.
The scalar curvature associated with this geometry has the form, 
\begin{equation}
R = - h''(r) - \frac{6}{r}h'(r)-\frac{6}{r^2}h(r)+\frac{6}{r^2}
\label{Eq:ricciscalar}
\end{equation}
As a consequence of imposing the constant-curvature conditions 
$R=A=\text{constant}$, Eq.\ (\ref{Eq:ricciscalar}) becomes a linear 
differential equation of the metric function $h(r)$. This substantial 
simplification reduces the original fourth-order field equations to 
a second-order problem that can be solved exactly. The solution 
to this linear differential equation is given by, 
\begin{equation}
h(r) = 1- \frac{A}{20}r^2+\frac{C_1}{r^2}+\frac{C_2}{r^3}, 
\label{Eq:gensolution}
\end{equation}
 where $C_1$ and $C_2$ are constants of integration. The resulting 
 solution characterizes the most general, spherically symmetric geometry 
 compatible with the constant-curvature Ansatz and thus provides 
 a natural starting point for investigating higher-dimensional black hole 
 solutions within the Starobinsky-type models. 

\subsection{Branch Structure of Solutions}
Solutions of Eq.\ (\ref{Eq:reducedFE}) are categorized into 
two branches. The first is the degenerate branch, where $f'(A) = 0$. 
The second, the non-pathological branch, occurs when $f'(A)\neq 0$, 
and by virtue of the ghost stability condition $f'(A)>0$, we only 
consider solutions in this branch satisfying this constraint. 

\subsubsection{Degenerate (Pathological) Branch \texorpdfstring{$f'(A)= 0$}{f'(A)= 0}}
In this case the field equations lose their dynamical character entirely. 
The Ricci tensor is no longer fixed by the metric, and the spacetime geometry is governed 
by the scalar condition alone,
\begin{equation}
    \La = -\frac{1}{8\alpha}, \,\,\, A = -\frac{1}{2\alpha},
\label{Eq:LaConditionDegen}
\end{equation}

providing us with the relation, 
\begin{equation}
    A = 4\La.
\end{equation}
Consequently, the metric is no longer uniquely determined and additional 
integration constants become permissible. Substituting this relation into 
Eq.\ (\ref{Eq:gensolution}), we obtain, 
\begin{equation}
    h(r) = 1-\frac{\La}{5}r^2+\frac{C_1}{r^2}+\frac{C_2}{r^3}.
\end{equation}
The presence of the additional $\mathcal{O}(1/r^3)$ contribution 
reflects the degeneracy of the field equations in this branch. Since the 
equations no longer constrain the spacetime geometry, these kinds of 
solutions are typically regarded as non-physical. 

\subsubsection{Non-pathological Branch \texorpdfstring{$f'(A) > 0$}{f'(A) > 0}}

In the non-pathological case,
\begin{equation}
A > -\frac{1}{2\alpha},
\end{equation}
and the field equations reduce to
\begin{equation}
    \R  = \frac{A}{5}\g.
\label{Eq:nonpatheq}
\end{equation}
As a consequence, the spacetime geometry is no longer determined 
just by the constant-curvature Ansatz, but must also satisfy the underlying 
tensor structure of the field equations. While the general solution in 
Eq.\ (\ref{Eq:gensolution}) admits both $\mathcal{O}(1/r^2)$ and 
$\mathcal{O}(1/r^3)$ contributions, direct substitution into 
Eq.\ (\ref{Eq:nonpatheq}) demonstrates that the $\mathcal{O}(1/r^3)$ 
term generates additional radial dependence within the Ricci tensor 
that is incompatible with the Einstein-space condition. Therefore, any 
contributions to the metric that introduce additional radial dependence 
incompatible with constant curvature must vanish, effectively eliminating the 
$\mathcal{O}(1/r^3)$. The metric function is thus reduced to
\begin{equation}
h(r) = 1-\frac{A}{20}r^2+\frac{C_1}{r^{2}},
\end{equation}
which is identified with the five-dimensional Tangherlini 
solution \cite{Tangherlini1963bw}. The constant of integration $C_1$ is 
identified with the Tangherlini mass parameter \cite{Emparan_2008}, and 
is thus proportional to the mass of the five-dimensional black hole. 
The curvature scale is determined by the effective cosmological constant,
\begin{equation}
\La_{\text{eff}} = \frac{A}{5}. 
\end{equation}
The quadratic curvature correction, therefore, does not introduce 
new independent hair in the ghost-free branch. Instead, its effect is encoded 
through the modified relation between the constant curvature $R=A\neq 0$, 
the bare cosmological constant $\La$, and the coupling parameter $\a$. 
In this branch, the higher-curvature terms act as a source of an emergent 
cosmological constant while preserving the standard Einstein-space structure of the vacuum.

As an additional check on the physical viability of this branch 
we consider the mass of the scalaron degree of freedom. Linearization of 
the trace equation around the constant curvature background $R =A $ 
gives the Klein-Gordon equation for curvature perturbation, 
\begin{equation}
(\Box - m_s^2)\delta R = 0,
\end{equation}
with the scalaron mass defined as
\begin{equation}
m_s^2 = \frac{\left(D-2\right)f'(A)-Af''(A)}{2(D-1)f''(A)}.
\end{equation}
In the case of the five-dimensional Starobinsky model, we have
\begin{equation}
m_s^2 = \frac{3+2\alpha A}{16\alpha}.
\end{equation}
Substitution of the constant curvature solution found in Eq.\ (\ref{Eq:Asol}) gives 
\begin{equation}
m_s^2 = \frac{\sqrt{9+40\alpha \La}}{16\alpha}.
\end{equation}
For $\alpha >0$, the scalaron condition, $m_s^2>0$ requires that $\La > -\frac{9}{40\alpha}$, a weaker requirement than that of the ghost-free condition $\La > -\frac{1}{8\alpha}$ already previously obtained. Thus, under the constant curvature analysis performed on this model, every solution on the ghost-free branch automatically satisfies the scalaron non-tachyonic condition. The implications of the scalaron mass bounds and their implications on the parameter space is beyond the scope of this work and is left for future investigation. 

\subsection{Presence of emergent vacua in the branch structure}
The presence of emergent vacua is characterized by solutions where $A\nrightarrow 0 $ as $\L \to 0$. In the $\La \to 0$ limit, the degenerate branch structure ceases to exist, as Eq.\ (\ref{Eq:LaConditionDegen}) imposes a condition that is incompatible with the limit. The non-pathological branch however, admits two solutions to Eq.\ (\ref{Eq:scalareqn}), 
\begin{equation}
A = 0, \quad \frac{-3}{\alpha}.
\end{equation}
These solutions reduce the positive branch, $f'(A)>0$ to a flat spacetime ($A=0$). The negative branch, although admitting a solution with $A\neq 0$ despite $\La \to 0$, violates the imposed ghost instability condition and must be discarded altogether. Therefore, the addition of a pure quadratic curvature correction, fails to generate emergent, ghost-free vacuum solutions in the absence of a bare cosmological constant. 

\section{Generalization to Arbitrary Dimensions and Higher-Order Curvature}
\label{sec:singleTermCorrection}

Having demonstrated that pure $R^2$ gravity fails to generate a stable, emergent vacuum in five dimensions without a bare cosmological constant, it is instructive to generalize this result. We now ask whether a dynamically generated, ghost-free vacuum is possible in arbitrary spacetime dimensions $D$ by considering generalized non-linear curvature corrections of the form,
\BEq
f(R) = R + \alpha R^n, \quad n > 1. 
\EEq
To isolate the purely geometric generation of vacuum energy, we set the 
bare cosmological constant to zero ($\La = 0$).

In an arbitrary dimensions $D$, taking the trace of the generalized $f(R)$ field 
equations,
\begin{equation}
\label{eq:f(R)-field-eqs}
f'(R) R_{\mu\nu}
- \frac{1}{2} f(R) g_{\mu\nu}
- \left[ \nabla_\mu \nabla_\nu - g_{\mu\nu} \Box \right] f'(R)
= 0, \quad \Box \equiv g^{\a\b}\nabla_{\a}\nabla_{\b},
\end{equation}
produces the scalar constraint,
\begin{equation}
f'(R) R - \frac{D}{2} f(R) + (D-1) \Box f'(R) = 0.
\end{equation}
Imposing the constant curvature Ansatz $R = A \neq 0$, the d'Alembertian 
term vanishes, reducing the field equations to the master algebraic constraint,
\begin{equation}
A f'(A) - \frac{D}{2} f(A) = 0.
\label{Eq:MasterConstraint}
\end{equation}
Substituting the generalized Lagrangian $f(A) = A + \alpha A^n$ into 
Eq.\ (\ref{Eq:MasterConstraint}), we obtain
\begin{equation}
A \left( 1 + n\alpha A^{n-1} \right) - \frac{D}{2} \left( A + \alpha A^n \right) = 0.
\end{equation}
Assuming a non-trivial emergent vacuum ($A \neq 0$), we divide by $A$ and 
rearrange to solve for the effective geometric coupling,
\begin{equation}
\alpha A^{n-1} = \frac{D - 2}{2n - D}.
\label{Eq:AlphaRoot}
\end{equation}
For this mathematically allowed emergent vacuum to be physically viable, 
it must satisfy the ghost-free stability criterion, requiring a positive 
effective gravitational coupling, $f'(A) > 0$. Evaluating the derivative of the 
action at the constant-curvature root and substituting the relation from 
Eq.\ (\ref{Eq:AlphaRoot}), we find,
\begin{equation}
f'(A) = 1 + n(\alpha A^{n-1}) = 1 + n \left( \frac{D - 2}{2n - D} \right).
\end{equation}
Gathering terms over a common denominator, the expression simplifies to
\begin{equation}
f'(A) = \frac{2n - D + nD - 2n}{2n - D} = \frac{D(n - 1)}{2n - D}.
\label{Eq:GeneralizedStability}
\end{equation}

This final relation, Eq.\ (\ref{Eq:GeneralizedStability}), reveals an interesting feature 
of emergent vacua in modified gravity: the physical stability of the dynamically 
generated vacuum is completely independent of the coupling strength $\alpha$. 
It is governed exclusively by the spacetime dimensionality $D$ and the polynomial 
order of the curvature correction $n$. 

Because we are considering macroscopic spacetimes ($D \geq 4$) and higher-order 
UV corrections ($n > 1$), the numerator $D(n - 1)$ is strictly positive. Consequently, 
the condition for a ghost-free vacuum ($f'(A) > 0$) demands that the denominator 
also be positive, yielding the universal stability bound,
\begin{equation}
n > \frac{D}{2}.
\label{Eq:UniversalBound}
\end{equation}

This universal bound provides immediate physical insight into the limitations of 
purely geometric vacuum generation. For a standard four-dimensional spacetime 
($D=4$), stability requires $n > 2$; hence, the classic $R^2$ Starobinsky model 
cannot spontaneously generate a stable de Sitter or anti-de Sitter vacuum without 
a bare cosmological constant. In five dimensions ($D=5$), the bound requires 
$n > 2.5$. This explains the failure of the five-dimensional Starobinsky model analyzed 
in \cref{sec:5dStarobinsky}: the $R^2$ correction is simply of insufficient polynomial order to stabilize 
the emergent geometry. Generating a stable emergent vacuum in five dimensions 
would require at least a cubic curvature correction ($n=3$). Furthermore, in the 
context of ten-dimensional superstring effective actions ($D=10$), an emergent 
vacuum would require at least $R^6$ corrections. Thus, Eq.\ (\ref{Eq:UniversalBound}) 
establishes a rigid, generalized ``no-go'' condition for the dynamical generation 
of stable geometric vacua in arbitrary dimensions.

\section{Emergent Vacua in Extended Polynomial Gravity}
\label{sec:multiTermCorrection}

While the strict bound $n > D/2$ derived previously establishes a firm limitation 
on single-term higher-curvature corrections, realistic effective field theories 
typically manifest as a full polynomial expansion of the Ricci scalar. In this section, 
we demonstrate that introducing a multi-term polynomial hierarchy provides 
the necessary additional degrees of freedom to bypass the single-term restriction, 
allowing for the dynamical generation of stable, ghost-free vacua in arbitrary 
dimensions $D \ge 5$ without a bare cosmological constant, $\La = 0$.

Let us consider a generalized gravitational action featuring a polynomial expansion 
up to order $n$:
\begin{equation}
    f(R) = R + \sum_{k=2}^n c_k R^k.
\end{equation}
Following the same procedure as before, we impose the constant scalar curvature 
Ansatz $R = A \neq 0$. The trace of the field equations reduces to the generalized 
algebraic constraint,
\begin{equation}
    A \lr{ 1 + \sum_{k=2}^n k c_k A^{k-1} } - \frac{D}{2} \lr{ A + \sum_{k=2}^n c_k A^k } = 0.
\end{equation}
Dividing by $A$ and grouping the summation terms, we can express this constraint 
linearly. It is convenient to define a set of dimensionless coupling 
parameters, $x_k \equiv c_k A^{k-1}$. The emergent vacuum constraint then takes 
the compact form,
\begin{equation}
    \sum_{k=2}^n (D - 2k) x_k = 2 - D.
    \label{Eq:PolynomialConstraint}
\end{equation}
Simultaneously, the physical viability of the dynamically generated vacuum relies 
on the absence of ghost instabilities, enforcing the stability condition on the 
effective gravitational coupling,
\begin{equation}
    f'(A) = 1 + \sum_{k=2}^n k x_k > 0.
    \label{Eq:PolynomialStability}
\end{equation}

When the expansion is truncated at $n=2$ (the pure Starobinsky model), 
the system contains only a single variable, $x_2$. Eq.\ (\ref{Eq:PolynomialConstraint}) 
fixes this parameter to $x_2 = (2-D)/(D-4)$, leaving no remaining degrees 
of freedom to satisfy the inequality in Eq.\ (\ref{Eq:PolynomialStability}). Consequently, 
$f'(A)$ is negative for all $D \ge 5$.

However, if we extend the theory to incorporate at least an $\mathcal{O}(R^3)$ 
correction ($n=3$), the system possesses two variables ($x_2$ and $x_3$) governed 
by a single linear constraint. This allows the lower-order term to satisfy the geometric 
trace condition while the higher-order term independently drives the effective 
coupling into the positive, ghost-free regime. To illustrate this mechanism, we 
analyze the 
\BEq
f(R) = R + c_2 R^2 + c_3 R^3
\EEq
model in $D=5$, $6$, and $7$ dimensions.

\subsection{Five Dimensions (\texorpdfstring{$D=5$}{D=5})}
In five dimensions, the trace constraint from Eq.\ (\ref{Eq:PolynomialConstraint}) gives
\begin{equation}
(5 - 4)x_2 + (5 - 6)x_3 = 2 - 5 \implies x_2 - x_3 = -3.
\end{equation}
Solving for $x_3$, we find $x_3 = x_2 + 3$. Substituting this into the stability condition gives
\begin{equation}
f'(A) = 1 + 2x_2 + 3x_3 = 1 + 2x_2 + 3(x_2 + 3) = 10 + 5x_2.
\end{equation}
For a stable vacuum ($f'(A) > 0$), we require $x_2 > -2$. Because $x_2$ is a 
free parameter of the theory, a wide parameter space exists for physically viable 
solutions. For example, if a theory lacks an $R^2$ term ($x_2 = 0$) but contains 
an $R^3$ term, the constraint fixes $x_3 = 3$, yielding a strongly positive effective 
coupling $f'(A) = 10$.

\subsection{Six Dimensions (\texorpdfstring{$D=6$}{D=6})}
Six dimensions presents a unique scenario. The constraint equation 
becomes
\begin{equation}
(6 - 4)x_2 + (6 - 6)x_3 = 2 - 6 \implies 2x_2 = -4.
\end{equation}
Remarkably, the coefficient for $x_3$ vanishes, and the trace condition  
fixes the quadratic coupling to $x_2 = -2$, while placing no restriction 
on the cubic coupling $x_3$. The stability bound evaluates to
\begin{equation}
f'(A) = 1 + 2(-2) + 3x_3 = 3x_3 - 3.
\end{equation}
To achieve stability, we must simply mandate $x_3 > 1$. Thus, in six dimensions, 
the $R^2$ term is exclusively responsible for sourcing the emergent curvature, 
while the $R^3$ term acts to stabilize the graviton kinetic energy. A choice 
of $x_3 = 2$, for instance, results in $f'(A) = 3 > 0$.

\subsection{Seven Dimensions (\texorpdfstring{$D=7$}{D=7})}
In seven dimensions, the interplay between the terms continues. 
The constraint is given by
\begin{equation}
(7 - 4)x_2 + (7 - 6)x_3 = 2 - 7 \implies 3x_2 + x_3 = -5.
\end{equation}
Substituting $x_3 = -5 - 3x_2$ into the effective coupling we get
\begin{equation}
f'(A) = 1 + 2x_2 + 3(-5 - 3x_2) = -14 - 7x_2.
\end{equation}
Enforcing $f'(A) > 0$ requires $-14 - 7x_2 > 0$, or $x_2 < -2$. If we consider 
a parameter choice of $x_2 = -3$, the geometric constraint requires $x_3 = 4$. 
This configuration the results in a stable effective coupling of $f'(A) = 7$. 

\subsection{Explicit Lagrangians}

To make these stable, emergent solutions concrete, we can translate the 
dimensionless parameters $x_k$ back into the physical coupling constants 
of the Lagrangian via 
\BEq
c_k = \frac{x_k}{A^{k-1}}. 
\EEq
The specific parameter choices utilized in our examples correspond to the 
following explicit $f(R)$ theories, each dynamically generating a stable, 
ghost-free vacuum of constant curvature $A$:

\begin{itemize}
\item[] \textbf{5D Model ($x_2=0, x_3=3$):} 
    \begin{equation}
        f(R) = R + \frac{3}{A^2} R^3
    \end{equation}
\item[] \textbf{6D Model ($x_2=-2, x_3=2$):}
    \begin{equation}
        f(R) = R - \frac{2}{A} R^2 + \frac{2}{A^2} R^3
    \end{equation}
\item[] \textbf{7D Model ($x_2=-3, x_3=4$):}
    \begin{equation}
        f(R) = R - \frac{3}{A} R^2 + \frac{4}{A^2} R^3
    \end{equation}
\end{itemize}
In each of these modified gravity theories, the specific hierarchy of the $R^2$ 
and $R^3$ couplings ensures that the trace constraint is satisfied while simultaneously 
satisfying the stability condition $f'(A) > 0$.

These examples explicitly demonstrate that while pure quadratic gravity inevitably 
generates pathological emergent vacua in $D \ge 5$, the inclusion of higher-order 
polynomial terms restores physical viability. A mixed curvature hierarchy extending 
to at least $\mathcal{O}(R^3)$ contains sufficient degrees of freedom to dynamically 
generate a stable, exact vacuum spacetime entirely from geometric corrections, 
independent of a fundamental cosmological constant.

\subsection{Global Stability Constraints}

The stability condition, $f'(A) > 0$, guarantees that 
the asymptotic background spacetime is ghost-free. However, a physically 
viable theory must remain stable across all dynamic curvature scales, particularly 
in localized regions of high curvature such as stellar interiors or near black hole 
singularities. Thus, we must extend our analysis to identify the subset of parameter 
space that guarantees \textit{global} stability, requiring $f'(R) > 0$ for all $R \in \mathbb{R}$.

Focusing on the $\mathcal{O}(R^3)$ model, the effective gravitational coupling 
evaluated at an arbitrary curvature $R$ is a quadratic function,
\begin{equation}
f'(R) = 1 + 2c_2 R + 3c_3 R^2.
\end{equation}
For this coupling to remain positive for all real values of $R$, the function 
must describe an upward-opening parabola with no real roots. This imposes two 
requirements: a positive leading coefficient ($c_3 > 0$) and a 
negative discriminant ($\Delta = b^2 - 4ac < 0$). Evaluating the discriminant 
we find the global stability condition for the physical couplings,
\begin{equation}
(2c_2)^2 - 4(3c_3)(1) < 0 \implies c_3 > \frac{1}{3} c_2^2.
\end{equation}
Multiplying this inequality by $A^2$ (which is positive for any non-trivial 
vacuum) maps this condition into our dimensionless parameter space,
\begin{equation}
\label{Eq:GlobalStability}
x_3 > \frac{1}{3} x_2^2.
\end{equation}
This establishes a rigorous scale-independent boundary: for 
an $\mathcal{O}(R^3)$ theory to be globally stable, its dimensionless couplings 
must lie in the region bounded above the parabola $x_3 = x_2^2/3$. By intersecting 
this global requirement with the linear trace constraints derived previously, we can 
sharply define the globally stable subset of theories for any arbitrary dimension $D$.\\

\textbf{Five Dimensions ($D=5$):} 
Substituting the 5D trace constraint ($x_3 = x_2 + 3$) into the global stability bound 
gives,
\begin{equation}
x_2 + 3 > \frac{1}{3} x_2^2 \implies x_2^2 - 3x_2 - 9 < 0.
\end{equation}
Solving for the roots of this quadratic defines the globally stable parameter interval,
\begin{equation}
\frac{3 - 3\sqrt{5}}{2} < x_2 < \frac{3 + 3\sqrt{5}}{2}.
\end{equation}
This restricts the quadratic coupling to approximately $x_2 \in (-1.85, 4.85)$. 
Our previous 5D example ($x_2 = 0$) falls comfortably within this bound, ensuring 
it is a globally viable theory.\\

\textbf{Six Dimensions ($D=6$):}
In 6D, the trace constraint locked the quadratic coupling to $x_2 = -2$. 
Substituting this value into the global bound gives
\begin{equation}
x_3 > \frac{1}{3} (-2)^2 \implies x_3 > \frac{4}{3}.
\end{equation}
While our initial local stability analysis merely required $x_3 > 1$, global stability 
enforces a tighter restriction. Our 6D example ($x_3 = 2$) safely satisfies this more 
rigorous $x_3 > 1.33$ condition.\\

\textbf{Seven Dimensions ($D=7$):}
Substituting the 7D trace constraint ($x_3 = -5 - 3x_2$) into the global bound produces
\begin{equation}
-5 - 3x_2 > \frac{1}{3} x_2^2 \implies x_2^2 + 9x_2 + 15 < 0.
\end{equation}
This restricts the 7D parameter space to the closed interval,
\begin{equation}
\frac{-9 - \sqrt{21}}{2} < x_2 < \frac{-9 + \sqrt{21}}{2}.
\end{equation}
Consequently, global stability in seven dimensions demands $x_2 \in (-6.79, -2.21)$. 
Our earlier choice of $x_2 = -3$ lies within this valid subset, verifying that the resulting 
model remains ghost-free across all curvature scales.

\section{Discussion}
In this work, we have systematically investigated the necessary conditions for the dynamical generation of stable, ghost-free vacua in higher-dimensional $f(R)$ gravity. By analyzing static, spherically symmetric spacetimes characterized by a constant scalar curvature ($R=A)$, we clarified the branch structure of the permitted solutions. Our initial analysis of the five-dimensional Starobinsky model demonstrated that pure $R^2$ corrections cannot spontaneously generate a stable, exact vacuum spacetime; an effective, positive gravitational coupling ($f'(A) > 0$) in this specific model requires 
the presence of a bare cosmological constant.

While the criteria used in this analysis are standard 
within the modified gravity literature, their synthesis results in an interesting theoretical 
perspective. Specifically, the algebraic trace constraint, $A f'(A) - \frac{D}{2} f(A) = 0$, 
and the Dolgov-Kawasaki stability condition, $f'(R) > 0$, are routinely employed to 
verify the viability of specific, pre-supposed cosmological or black hole solutions. 
In contrast, by combining these two requirements into an \textit{a priori} algebraic 
approach, we have mapped the physically viable parameter space of the underlying 
theories themselves across arbitrary dimensions. This led to the derivation 
of a universal dimensional bound: for any single-term geometric extension 
of the form $f(R) = R + \alpha R^n$ 
to generate a stable emergent vacuum without a bare cosmological constant, 
the polynomial order must satisfy $n > D/2$. This result provides a rigorous 
explanation for why intuitions developed in four-dimensional modified gravity 
often fail in higher dimensions. While the classic $R^2$ Starobinsky model is 
sufficient to drive viable geometric dynamics in four dimensions, 
our bound shows that stable vacuum generation in five dimensions requires 
higher-order invariants, severely restricting the landscape of permissible 
higher-dimensional single-term models.

However, the theoretical value of this ``no-go'' bound is fully realized when 
placed in the context of effective field theory (EFT). We demonstrated that 
treating the higher-curvature corrections as a multi-term polynomial hierarchy 
extending to at least $\mathcal{O}(R^3)$ bypasses the single-term restriction. 
By reducing the non-linear differential field equations to a system of linear algebraic 
inequalities, we provided a precise methodology for constructing vacua in any 
dimension $D \ge 5$. We then extended this approach beyond 
asymptotic stability at the vacuum ($R=A$) to ensure global stability, requiring 
$f'(R) > 0$ for all real values of $R$. This analysis reveals a scale-independent 
cooperative dynamic between curvature powers: lower-order polynomial terms 
can be constrained to satisfy the geometric trace condition, while higher-order 
terms possess the necessary degrees of freedom to drive the effective gravitational 
coupling into the globally stable, positive regime. The explicit parameter boundaries 
derived for five, six, and seven dimensions illustrate that higher-order curvature terms 
are not merely phenomenological add-ons, but necessary components 
for stabilizing emergent geometries across all dynamic curvature scales.

Ultimately, these results clarify the fundamental relationship between spacetime 
dimensionality, geometric degrees of freedom, and vacuum stability. Future work 
could naturally extend this algebraic methodology to evaluate the thermodynamic 
stability of the resulting black hole solutions, or explore how the inclusion of explicit 
matter fields modifies the boundaries of the globally stable parameter space. Furthermore, 
applying a similar algebraic synthesis to more complex curvature invariants, such as those 
found in generic Lovelock or Gauss-Bonnet theories, may reveal analogous dimensional 
bounds governing the global stability of higher-order gravitational models.

\bibliographystyle{apsrev4-2}
\bibliography{References_VFinal}

\end{document}